# Scanned Probe Microscopy of Electronic Transport in Carbon Nanotubes

A. Bachtold[1,2], M. S. Fuhrer[1,2], S. Plyasunov[1,2], M. Forero[1,2], Erik H. Anderson[2], A. Zettl[1,2] and Paul L. McEuen[1,2]

[1]*Department of Physics, University of California, Berkeley, CA 94720,*
[2]*Materials Sciences Division, Lawrence Berkeley National Laboratory, Berkeley, CA 94720*

(February 12, 2000)

We use electrostatic force microscopy and scanned gate microscopy to probe the conducting properties of carbon nanotubes at room temperature. Multi-walled carbon nanotubes are shown to be diffusive conductors, while metallic single-walled carbon nanotubes are ballistic conductors over micron lengths. Semiconducting single-walled carbon nanotubes are shown to have a series of large barriers to conduction along their length. These measurements are also used to probe the contact resistance and locate breaks in carbon nanotube circuits.

Electronic devices constructed from single-walled carbon nanotubes (SWNTs) and multiwalled carbon nanotubes (MWNTs) show remarkable behavior. Individual SWNTs can act as conducting wires[1,2], field-effect transistors[3], or single-electron-tunneling transistors[1,2]. Combinations of nanotubes can act as rectifiers[4] or more complex multi-terminal devices[5]. Mutli-walled nanotubes have shown the Aharonov-Bohm effect[6] and have been used to construct spin-electronic devices[7].

Most of what is known about the transport properties of nanotubes comes from DC electrical measurements of single nanotubes or bundles. These transport experiments suggest that metallic SWNTs are remarkably good conductors, with very long mean free paths[1,2,8]. Doped semiconducting SWNTs, on the other hand, have much higher resistances, and recent experiments suggest that transport is not simply diffusive but instead limited by a series of large barriers along the nanotube length[9]. MWNTs have been shown to be diffusive conductors[6,10,11], although experiments on individual MWNTs contacted with liquid metal have been interpreted as evidence of ballistic conduction[12].

A complication for interpreting these experiments is the fact that the measured two-terminal transport characteristics do not uniquely identify the underlying device behavior. Using the Büttiker-Landauer formalism[13] in the ohmic (classical transmission) limit, the two-terminal resistance of nanotube is: $R = h/4e^2 + R_i + R_{c1} + R_{c2}$, where $h/4e^2$ is the (quantized) contact resistance of the nanotube. The additional contributions arise from physically separate mechanisms – the intrinsic resistance $R_i$ from scattering processes within the tube arising from, e.g. disorder or phonons, and the contact resistances $R_{c1,2}$ from the transport barriers formed at the metal electrode/nanotube junctions. Since the measured two-terminal resistance $R$ is the sum of these contributions, it is unclear from previous experiments the degree to which each of these contributions affects the measured conductance. Four terminal measurements can in principle alleviate the effects of the contacts, but typical macroscopic electrodes act as invasive probes that influence the resistance of the object studied[13].

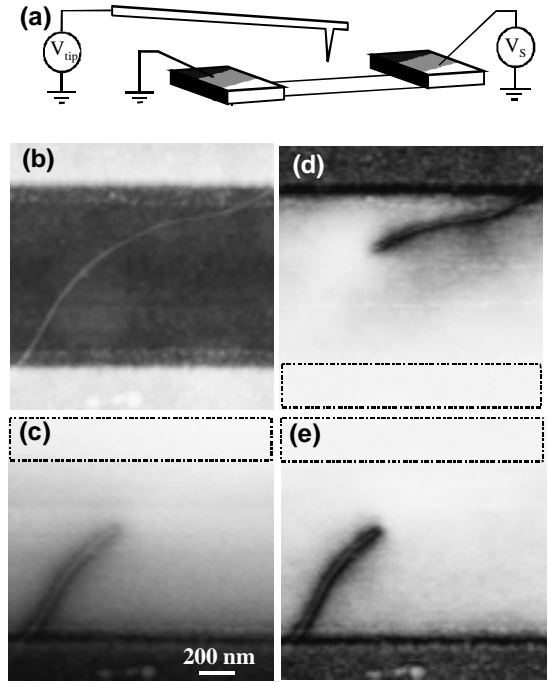

FIG. 1. (a) Experimental setup for EFM and SGM. A conducting AFM cantilever is scanned above the device, which consists of a nanotube contacted by two gold electrodes. The device is fabricated on a highly doped oxidized Si wafer, which is used as a gate electrode. (b) Topographic AFM image of a 2.5nm diameter bundle of SWNTs which has been broken electrically by the application of a large voltage (~6V). (c) AC-EFM image of the device where an AC potential of 100mV is applied to the lower electrode. (d) and (e) DC-EFM images where a DC potential of 1V is applied to the upper or lower electrode, respectively.

Here we use electrostatic force microscopy (EFM)[14] and scanned gate microscopy (SGM)[15] to directly probe the nature of conduction in SWNTs and

MWNTs. By using an AFM tip as a local voltmeter (EFM), we separately measure the intrinsic resistance and contact resistances of SWNTs and MWNTs. We show for the first time that in metallic SWNTs the measured resistance is due to contact resistance, i.e. electron transport is ballistic. Furthermore, by using the AFM tip as a local gate (SGM), we directly image individual scattering sites in semiconducting SWNTs and show that a series of large barriers limit transport.

We begin by reviewing the experimental techniques, starting with EFM (Fig. 1(a)). An AFM tip with a voltage $V_{tip}$ is scanned over a nanotube sample and the electrostatic force between the sample and tip is measured in one of two ways. In the first, called DC-EFM, the AFM is operated in non-contact mode with the cantilever oscillated near its resonant frequency at a small fixed height above the sample[16]. The changing electrostatic force with $z$ gives a shift in the resonant frequency and the phase $\Delta\varphi$ of the cantilever oscillation, $\Delta\varphi \propto (d^2C/dz^2)(V_{tip} + \phi - V_s)^2$, where $V_s$ is the voltage within the sample, $\phi$ is the work function difference between the tip and sample, $V_{tip}$ is the tip voltage, and $C$ is the tip-sample capacitance. The measured signal is thus proportional to the square of the DC voltage difference between the tip and the sample. In the second approach, called AC-EFM, the cantilever is made to oscillate by an AC potential that is applied to the sample at the resonant frequency of the cantilever. This produces an AC force on the cantilever proportional to the local AC potential $V_s(\omega)$ beneath the tip, $F_{ac}(\omega)=(dC/dz)(V_{tip}+\phi)V_s(\omega)$. The resulting oscillation amplitude is recorded using an external lock-in amplifier; the signal is proportional to $V_s(\omega)$[17].

Note that in both cases above there is a large potential difference $V_{tip} + \phi - V_s$ between the tip and the sample. The tip may therefore locally modify the conducting properties of the sample as it scans over it. Scanned gate microscopy (SGM) images this perturbation by measuring the conductance of the sample as a function of tip position. The conductance changes when the tip locally depletes, or gates the underlying electron system. This technique has been used in the past to study quantum point contacts[15] and quantum Hall conductors[18]. SGM may also be used to determine whether the tip perturbs the sample during an EFM measurement.

Having introduced the techniques, we now discuss measurements that clearly illustrate the utility of EFM. Fig. 1(b) is an AFM image of the topography of a nanotube device. It consists of a small bundle of SWNTs approximately 2.5 nm in diameter, likely consisting of a few 1.4 nm diameter tubes[19]. A large bias (~6V) was applied across the sample until it failed and the resistance became immeasurably large[20]. Subsequent topographic imaging showed no clear point of failure of the nanotube. EFM imaging, however, clearly reveals the location of the break, as shown below.

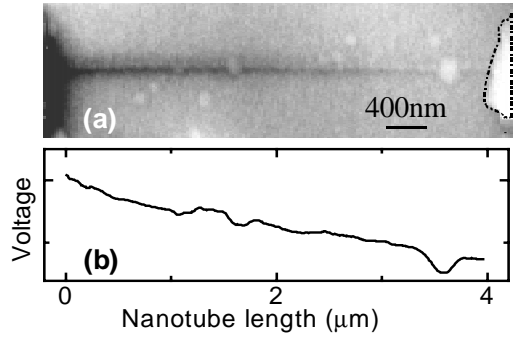

FIG. 2. (a) AC-EFM image of a MWNT of diameter 9nm. The resistance of the entire circuit is 42kΩ. An AC bias of 150mV is applied to the left electrode; the IV characteristic verified that this bias was within linear response. (b) AC-EFM signal as a function of the nanotube length.

Fig. 1(c) shows an AC-EFM image of the tube with an AC voltage of 100 mV applied to the lower contact while the upper contact is grounded. A strong signal is seen on the lower half of the tube while the upper half of the tube yields no signal. Fig. 1(e) shows a DC-EFM image of the same device with a DC voltage of 1V applied to the lower contact. The result is the same. Fig. 1(d) shows the DC-EFM image with the potential applied to the upper contact. Now the upper half of the tube is visible. The location of the break is thus clearly determined.

In this case, both DC-EFM and AC-EFM produce satisfactory images. However, each technique has specific strengths. DC-EFM produces a more local signal, since it is proportional to a higher derivative of the capacitance with respect to $z$. AC-EFM can be performed at much lower voltages ($V_{ac} \sim 100\ mV$) than DC-EFM ($V_{dc} \sim 1\ V$), allowing measurements in linear response. However, because of the less local nature of the AC-EFM measurement, there is a background signal due to stray capacitive coupling of the tip to the large metal electrodes. This background can be seen as a gradient in the signal in Fig. 1(c).

We have developed a technique for subtraction of the background signal from the image. We first select a portion of the image between the electrodes that is far from the nanotube. This portion of the signal is then fit with a simple polynomial function, which is then subtracted from the signal over the entire image. The result is a "flattened" image which, at least in the area between the electrodes, represents the signal due to the nanotube only. We have applied this technique in instances where a nanotube is connected to only a single electrode and the potential in the tube is expected to be constant (e.g. Fig. 1(c)). The procedure accurately reproduced the expected flat voltage profile for the nanotube. The AC-

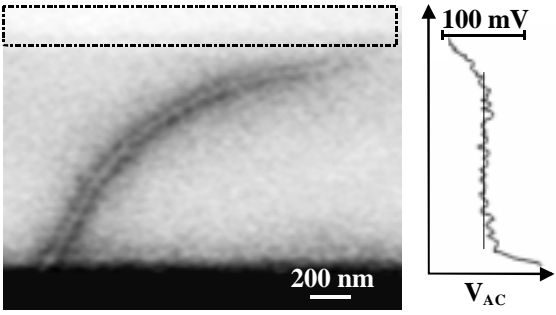

FIG. 3. An AC-EFM image of the same bundle of SWNTs shown in Fig. 1 before breaking. The bundle is metallic, with a total circuit resistance of 40kΩ. An AC potential of 100mV is applied to the lower electrode. The AC-EFM signal is flat along the length of the SWNT bundle. A trace of the potential as a function of vertical position in the image is also shown.

EFM images in Fig. 2-4 are shown with the background signal subtracted according to this procedure.

Fig. 2(a) shows an AC-EFM measurement of a typical MWNT with a diameter of 9nm and a two-terminal resistance of 42kΩ. The AC voltage is applied to left contact while the right contact is grounded. The AC-EFM signal drops uniformly along the MWNT length, with no significant drop at the contacts. This is verified in a line scan of the voltage along the tube length, shown in Fig. 2(b). The linear voltage drop indicates that the tube behaves as a diffusive conductor with a well-defined resistance per unit length, $R_i/L \sim 10 k\Omega/\mu m$. This confirms the results from previous transport[11] and scanned contact[21] measurements of MWNTs. We note that scanned gate measurements showed no appreciable signal on MWNTs (<1%), indicating that the tip did not significantly perturb the conducting properties of the sample.

We now turn to metallic SWNTs. We first discuss measurements of the device shown in Fig. 1 before electrical failure. The resistance of this 2.5 nm diameter bundle is 40 kΩ and has no significant gate voltage dependence - current is carried by metallic tubes in the bundle. At large biases the current saturates at ~ 50 µA. This is in agreement with recent work by Yao, Kane and Dekker[22] where the current was observed to be limited to 25 µA per metallic nanotube due to optical or zone-boundary phonon scattering. We therefore conclude that the current is carried by 2 metallic SWNTs in the bundle.

Fig. 3 shows the EFM image of this SWNT bundle, as well as a line trace along the backbone of the bundle. The potential is flat over its length, indicating that within our measurement accuracy there is no measurable intrinsic resistance. Taking into account the finite measurement resolution and possible errors introduced by the background subtraction, we estimate that $R_i$ is at most 3 kΩ. SGM showed no measurable effects, indicating that the tip did not perturb the conducting properties.

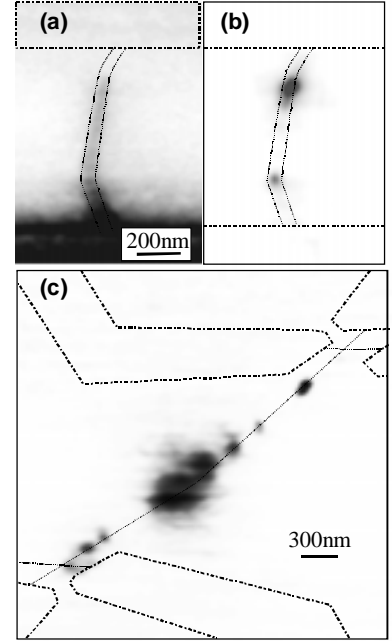

FIG. 4. (a) AC-EFM image of a 3nm diameter semiconducting bundle of SWNTs (lines indicate the position of the bundle). The device resistance is approximately 60MΩ. An AC potential of 100mV is applied to the lower electrode, while the tip is held at a DC bias of –2V. (b) SGM image for $V_{tip}$= +1V; black corresponds to immeasurably large resistance. (c) SGM image of a longer semiconducting SWNT bundle incorporated in a nanotube circuit (lines indicate the position of the bundle and the electrodes). Current is passed through the circuit from the upper left electrode to the lower right electrode.

We can relate the measured resistance to a transmission probability using the four-terminal Landauer formula: $R_i = (h/4e^2)(1-T_i)/T_i$ per nanotube, where $T_i$ is the transmission coefficient for electrons along the length of the nanotube. We find that $T_i > ½$ - the majority of electrons traverse the bundle without scattering. This unambiguously demonstrates that transport in metallic nanotubes is ballistic over a length of > 1µm, even at room temperature.

This result is consistent with theoretical predictions of very weak scattering in metallic SWNTs[9,23]. It also agrees with previous low temperature transport measurements indicating that long metallic SWNTs may behave as single quantum dots[1,9] and with room temperature measurements that exhibit low two-terminal resistance[8]. Here, however, we directly determine that the contact resistance is the dominant portion of the overall resistance. The contact resistance per tube is given by $R_{c1,2} = (h/4e^2) (1- T_{1,2})/T_{1,2}$, indicating that the transmission coefficients for entering and leaving the bundle are ~ 0.1 in this device. This is consistent with low temperature measurements where similar contacts to SWNTs have been shown to lead to Coulomb blockade[1,2] and have been exploited to observe Luttinger liquid tunneling behavior in nanotube devices[24].

We now turn to measurements of semiconducting SWNTs. Fig. 4(a) shows an AC-EFM image of a device. The ~3 nm diameter bundle (whose outline is indicated by the dotted lines) is bent at two locations, defining three distinct segments. The resistance of the bundle is ~ 60 MΩ and is strongly dependent on the voltage applied on the back-gate, indicating a semiconducting tube. The EFM image (Fig. 4(a)) shows that the voltage drops occur along the nanotube length, particularly at the junctions between segments.

Unlike for the MWNTs and metallic SWNTs, scanned gate microscopy showed very strong effects for this semiconducting SWNT bundle, as shown in Fig. 4 (b). The tip has the most dramatic effect on the conductance at the same places that the voltage drops in Fig. 4(a). In this case, the locations of strong gate effects correspond to bends in the device, but this is not a general feature. Fig. 4 (c) shows SGM of a longer semiconducting SWNT bundle without such dramatic structural bends. The tip is again observed to have large effects at particular sites along the bundle separated by ~ 100-400 nm.

These experiments directly show that the resistance of semiconducting SWNTs is dominated by a few strong barriers spaced by ~ 100 nm along the length. These points likely correspond to places where the local electron density is minimal and/or strong tunnel barriers exist. Such behavior was postulated based on previous transport measurements of the Coulomb blockade in semiconducting nanotubes[9], but is directly observed here for the first time. The microscopic origin of these scattering sites is not known, but could correspond to localized defects in the tube or to long-range electrostatic potential fluctuations associated with localized charges or surface contaminants[25]. Further EFM studies are underway to investigate the nature of these scattering sites.

In conclusion, we have used scanned probe microscopy to measure the properties of current-carrying nanotube circuits at room temperature. We can separately determine the contact resistances and intrinsic nanotube resistances, as well as locate breaks in non-conducting circuits. MWNTs are shown to be diffusive conductors, while metallic SWNTs are shown to be ballistic over micron lengths. EFM and SGM show that conduction in semiconducting SWNTs is dominated by a series of barriers along their length. These measurements provide a clear and detailed picture of electrical transport in nanotube devices. Furthermore, they demonstrate that EFM and SGM are powerful tools for characterizing the local properties of nanotube circuits. Such local information will become increasingly important as more complex multiple-nanotube circuits are investigated.

We acknowledge J. Esteve, S. Tans, C. Vale, and M. Woodside for contributions. We thank A. Rinzler and R. E. Smalley for supplying the nanotube materials. This work was supported by DOE (Basic Energy Sciences, Materials Sciences Division, the $sp^2$ Materials Initiative) and by DARPA (Moletronics Initiative). E.A. was partially supported by DARPA (Advanced Lithography Program under DOE contract number DE-AC03-76SF0009).